\documentstyle[fleqn]{article}
\title{\bf The duality relation between Glauber
 dynamics and the diffusion-annihilation
 model as a similarity
 transformation}
\author{J. E. Santos\\
{\it Department of Physics - University of Oxford}\\
{\it Theoretical Physics, 1 Keble Road,
 Oxford OX1 3NP, UK}
\\email address: jesantos@thphys.ox.ac.uk\\
 PACS numbers: 05.40.+j, 75.10.Hk, 02.50.Ey}
\date{}
\newcommand{\beq}{\begin{equation}}
\newcommand{\eeq}{\end{equation}}
\newcommand{\ket}[1]{\mbox{$ \mid #1\, \rangle$}}
\newcommand{\bra}[1]{\mbox{$ \langle\, #1\mid$}}
\newcommand{\noc}[1]{\mbox{$\hat{n}_{#1}$}}
\newcommand{\spup}[1]{\mbox{$\hat{s}^{+}_{#1}$}}
\newcommand{\spdo}[1]{\mbox{$\hat{s}^{-}_{#1}$}}
\newcommand{\spupdo}[1]{\mbox{$\hat{s}^{\pm}_{#1}$}}
\newcommand{\crea}[2]{\mbox{$\hat{#1}^{\dagger}_{#2}$}}
\newcommand{\anni}[2]{\mbox{$\hat{#1}_{#2}$}}

\newcommand{\etali}[1]{\mbox{
\normalsize $\eta_{\mbox{\scriptsize{$#1$}}}$}}
\newcommand{\spx}[1]{\mbox{$\hat{\sigma}^{x}_{#1}$}}
\newcommand{\spz}[1]{\mbox{$\hat{\sigma}^{z}_{#1}$}}
\newcommand{\spy}[1]{\mbox{$\hat{\sigma}^{y}_{#1}$}}
\newcommand{\undn}{\mbox{$\underline{n}$}}

\begin{document}
\maketitle
\begin{abstract}
 In this paper we address the
 relationship between
 zero temperature Glauber
 dynamics and the 
 diffusion-annihilation
 problem in the free fermion case.
 We show that the well-known
 duality transformation
 between the two problems
 can be formulated 
 as a similarity transformation
 if one uses appropriate
 (toroidal) boundary conditions.
 This allow us to establish and
 clarify the precise nature of the
 relationship between the two models. 
 In this way we obtain a
 one-to-one correspondence
 between observables and initial
 states in the two problems.
 A random initial state
 in Glauber dynamics is related 
 to a short range correlated state
 in the annihilation problem.
 In particular the long-time
 behaviour of the density in
 this state is seen to depend
 on the initial conditions.
 Hence, we show
 that the presence of correlations
 in the initial
 state determine
 the dependence  
 of the long time
 behaviour of the density
 on the initial conditions,
 even if such correlations 
 are short-ranged. We also
 apply a field-theoretical method
 to the calculation of
 multi-time correlation functions in
 this initial state.
\end{abstract}
\pagebreak
\section{Introduction}
 It has been known for a long time that 
 there is a relation between 
 Glauber dynamics 
 \cite{Glauber} and
 the symmetric diffusion
 problem in the presence of
 annihilation and deposition 
 of pairs of particles
 for a certain choice of
 the diffusion constant \cite{Racz},
 which corresponds 
 to the case in which this problem
 can be solved using free
 fermions \cite{Alcaraz,Schutz1}.
 Using this relation, Family
 and Amar \cite{Family}
 have computed the time evolution
 of the particle density
 in the transformed state of the
 annihilation problem
 that corresponds to random initial 
 conditions in Glauber
 dynamics at $\mbox{T}=0$. They have
 shown that their result only agrees with 
 the previously known results
 by Spouge \cite{Spouge} if one starts 
 with zero initial magnetization in the
 Glauber problem. In all other cases 
 the long time behaviour
 of the density shows a dependence on the 
 initial conditions, a result
 which differs from the well known universal 
 behaviour valid for random initial
 states in the annihilation problem
 \cite{Torney}. 
 This rises the question of
 the correspondence between initial states 
 in the Glauber and 
 annihilation problems and more generally 
 the relation between
 observables in the two systems. In this 
 paper we show that the 
 duality transformation between the two 
 systems is really a similarity
 transformation, if one uses sector
 dependent, toroidal boundary
 conditions in the Glauber model (see below). 
 We show
 that a random initial state in Glauber 
 dynamics is mapped through this
 similarity transformation to a state with 
 nearest neighbour
 correlations. Such state is translationally 
 invariant, which allow
 us to recover the  result by Family and Amar and 
 also to compute higher order
 correlators using a
 field-theoretic technique. The relation
 between these quantities in the transformed
 state and their counterparts in the random 
 initial state with density
 $1/2$ is explored and some calculations 
 are explicitly done. Finally,
 we emphasize the role of the
 correlations in the initial state for 
 the failure of the universality
 hypothesis.

 The structure of this paper is as follows: 
 in section 2, we present 
 Glauber dynamics in terms of a quantum
 spin chain and the definition 
 of a Temperley-Lieb algebra
 in terms of the spin $1/2$ operators. The
 associated Hecke algebra
 permits the construction of the similarity
 transformation to the
 reaction-diffusion model and also determines
 the boundary
 conditions of the system.  The
 correspondence between observables
 in Glauber and annihilation
 dynamics is also addressed.
 In section 3, we discuss the relation
 between the initial states in the two
 problems. In particular,
 we study the mapping between the
 random initial state in Glauber
 dynamics and a
 short-range correlated state
 in annihilation dynamics.
 Using these results and  those
 of section 2 we establish a correspondence
 between correlation functions
 in the two problems.  
 In section 4, we show that the
 short range correlated
 state is a translationally
 invariant state and  we show how
 to calculate multiple time correlation
 functions in this state using a
 field-theoretic technique. The
 density is explicitly computed and
 shown to agree with Family and Amar's
 result. We study the relation between
 two-point correlation functions
 in these states
 and the random initial state with
 density $1/2$ and recover the zero time 
 correlations as a special case. Finally, in
 section 5, we present our conclusions.

 For simplicity we discuss only the case of
 zero temperature Glauber dynamics although our
 results are easily generalised to other 
 problems with little modification,
 e.g. Glauber dynamics at finite temperature
 or the model of generalised dynamics considered
 by Peschel and Emery \cite{Peschel}, which
 maps by a duality transformation
 to a model of diffusing particles
 with pair annihilation and creation away from
 the free fermion line.

\section{The transformation law for the operators}
 It is well known \cite{Alcaraz,Alcaraz2}
 that certain reaction-diffusion systems provide
 physical realisations of Hecke algebras.
 The time evolution of these processes
 is described by a  master-equation 
 which can be conveniently written using an operator formalism,
 in which one assigns to each configuration of the 
 system a state vector in an Hilbert space \cite{Kadanoff}.
 The probability distribution is then represented by 
 a state vector obeying a ``Schr\"odinger'' equation
 $\partial_t\,\ket{\Psi}\;=\;-H\ket{\Psi}$ with
 $\ket{\Psi}=\sum_{\undn}\,P(\undn,t)\ket{\undn}$
 where $P(\undn,t)$ is the probability of
 finding con\-fi\-gu\-ra\-tion $\undn$ at time $t$
 and the set of different $\ket{\undn}$ is 
 supposed to be orthonormal and complete.
 The operator $H$ is a linear and in general
 non-hermitian operator encoding the rules of
 the stochastic process. For some systems
 of interest this operator can be written \cite{Alcaraz,Alcaraz2}
 as a sum of generators of Hecke algebras.
 Given that there exists a relation between
 reaction-diffusion systems and Glauber dynamics
 \cite{Racz} it is natural to ask if the evolution
 operator for this system can also be written in terms
 of Hecke algebra generators and if so, what conclusions
 can one draw from it. In order to show this
 we define the following operators
 on a lattice of $L$ sites
\begin{eqnarray}
 H^{\pm}&\equiv&\sum_{j=1}^{L}\,(1\,-
\,e_{2j-1}\,)\,(1\,-\,
(\,e_{2j}\,+\,e_{2j-2}\,-\,1\,)\,)\nonumber \\
 e_{2j-1}&=&\frac{1}{2}\,(1\,+\,\spx{j}\,)
\;\;\;1\leq j\leq L\nonumber \\
  e_{2j}&=&\frac{1}{2}\,(1\,+\,\spz{j}\,
\spz{j+1}\,) \;\;\;1\leq j\leq L-1
\label{1}
\end{eqnarray}
 and $e_{0}\;=\;e_{2L}$ which
 will be given below. The
 operators $\spz{j}$, $\spx{j}$
 are Pauli spin matrices at site $j$.

 The set $e_{j}\; (1\leq j\leq 2L-1)$
 forms a Temperley-Lieb algebra
 \cite{Temperley}, characterized
 by the relations
\begin{equation}
\left.
\begin{array}{ll}
 e_{j}^{2}\;=\;e_{j}\\ 
 e_{j}\,e_{j\pm 1}\,e_{j}\;=
\;\frac{1}{2}\,e_{j}\\
 e_{j}\,e_{i}\;=\;e_{i}\,e_{j}&
\mbox{if}\;\;\; \mid j-i \mid\;
\geq\; 2\enspace. \\
\end{array}
\right.
\label{2}
\end{equation}
 
 In order to construct 
 $e_{2L}$ explicitly we define
 the set of operators
 $g_{j}\; (1\leq j\leq 2L-1)$
 and the duality operator 
 $D$ by \cite{Levy}
\begin{eqnarray}
 g_{j}&=&(1+\mbox{i})
\,e_{j}\,-\,1 \nonumber\\
 D&=&\left(\prod_{j=1}^{2L-1}
\,g_{j}\right)\,X
\label{3}
\end{eqnarray}
 where $\prod_{i=1}^{2L-1}\,g_{j}$
 is the ordered product of the $g_{j}$'s
 and $X$ will be either $1$ or $\spz{L}$
 in this paper. The important
 points about these
 two operators are that both
 commute with $g_{j}$ for
 $1\leq j\leq 2L-2$ and that $(g_{2L-1}\,X)^{2}
 =(X\,g_{2L-1})^{2}$.
 When $X=1$, we will call the
 corresponding duality operator $D_{+}$
 and when $X=\spz{L}$,
 we will call it $D_{-}$.
 As we will see below the choice 
 of the operator $D$ is directly
 related with the different types
 of toroidal boundary conditions that 
 were referred in the last section. 
 The operators $g_{j}$, together
 with the commutation relations
 with $X$, form an affine Hecke
 algebra associated with the
 Temperley-Lieb algebra given above
 (see \cite{Alcaraz,Alcaraz2,Levy} and
 references therein) and one finds
 $D\,e_{j}\;=\;e_{j+1}\,D$ for
 $1\leq j\leq 2L-2$. Since
 $g_{j}^{\dagger}\;=\;(1-\mbox{i})
 \,e_{j}-1$, one finds from 
 (\ref{3}) that $g_{j}^{\dagger}
 \,g_{j}\;=\;1$. Also $X^{2}=1$, and
 we conclude that $D$ is unitary and hence
 invertible. We define $e_{2L}$ as
\begin{equation}
 e_{2L}\;=\;D\,e_{2L-1}\,D^{-1}\enspace.
\label{4}
\end{equation}

 The set of operators $e_{j}\;
 (1\leq j\leq 2L-1)$ and $e_{2L}$
 satisfies the relations of a periodic
 Temperley-Lieb algebra \cite{Levy}
 with $2L$ generators which is
 defined by (\ref{2}) together with
 similar relations for $e_{2L}$
 ($e_{2L+1}=e_{1}$, etc). For the
 particular choices of $X$ given
 above, it can be shown that \cite{Schutz2}
\begin{equation}
\left\{
\begin{array}{ll}
 e_{2L}\;=\;\frac{1}{2}\,(\,1\,+
 \,\hat{C}\,\spz{L}\,\spz{1}\,)
 & \mbox{if}\;\;\; X=1\\
 e_{2L}\;=\;\frac{1}{2}\,(\,1
 \,-\,\hat{C}\,\spz{L}\,\spz{1}\,)
 & \mbox{if}\;\;\; X=\spz{L}\\
\end{array}
\right.
\label{5}
\end{equation}
 where $\hat{C}\;=\;\prod_{j=1}^{L}
\spx{j}$. If we substitute the
 definitions of the $e_{j}$'s in
 (\ref{1}) $H^{\pm}$
 can be seen to be the 
 generator of the time evolution
 for Glauber dynamics.
 The $+$ ($-$) sign stands
 when $X=1$ ($X=\spz{L}$). This
 explains the use of the notation
 $H^{\pm}$. The 
 choice $X=1$ ($X=\spz{L}$),
 corresponds to the operator
 $H^{+}$ ($H^{-}$)
 which, when applied to
 eigenstates of $\hat{C}$
 with eigenvalue
 $1$ ($-1$), generates the time evolution
 for Glauber dynamics
 at $\mbox{T}=0$  with 
 periodic boundary conditions 
 \cite{Henkel}. When applied
 to eigenstates of $\hat{C}$ with
 eigenvalue $-1$ ($1$)
 $H^{+}$ ($H^{-}$)
 generates the time evolution for Glauber dynamics
 with anti-periodic boundary conditions.
 Both dynamics are stochastic.
 We see that one can indeed
 write the evolution operator
 for Glauber dynamics in terms of
 Hecke algebra generators.
 We now define the
 following similarity transformation
\beq
 V_{\pm}\;=\;R\,D_{\pm}
\label{7}
\eeq
 where $R\;=\;\exp\left(\mbox{i}\frac{\pi}{4}
 \sum_{j=1}^{L}\spy{j}\right)$
 is a global rotation of $\frac{\pi}{2}$
 around the $y$ axis.
 From the relations between the
 Temperley-Lieb generators $e_{j+1}\;
 =\;D\,e_{j}\,D^{-1}$ it can
 be easily shown that
\begin{equation}
\left.
\begin{array}{ll}
 V_{\pm}\,\spx{j}\,V_{\pm}^{-1}\;=\;
\spx{j}\,\spx{j+1}& 1\leq j\leq L-1\\ 
 V_{\pm}\,\spz{j}\,\spz{j+1}
\,V_{\pm}^{-1}\;=\;
\spz{j+1}& 1\leq j\leq L-1\nonumber\\ 
 V_{\pm}\,\spx{L}\,V_{\pm}^{-1}\;=\;
\pm\,\hat{Q}_{L}\,\spx{L}\,\spx{1}\\ 
 V_{\pm}\,\hat{C}\,\spz{L}\,\spz{1}
\,V_{\pm}^{-1}\;=\;
\pm\,\spz{1}
\end{array}
\right.
\label{8}
\end{equation}
 where $\hat{Q}_{L}\;=
\;R\,\hat{C}\,R^{-1}\;=\;
\prod_{j=1}^{L}\spz{j}$. 
 From (\ref{8}) it also
 follows that $V_{\pm}\,
 \hat{C}\,V_{\pm}^{-1}\;=\;
 \pm\,\hat{Q}_{L}$. 
 Applying $V_{+}$ to $H^{+}$ and
 $V_{-}$ to $H^{-}$ gives 
\begin{eqnarray}
\tilde{H}^{\pm}&=&V_{\pm}
\,H^{\pm}\,V_{\pm}^{-1}
\;=\;\sum_{j=2}^{L-1}\,\frac{1}{2}\,
 (1\,-\,\spx{j}\,\spx{j+1}\,)
 \,(1-\frac{1}{2}\,
 (\,\spz{j}\,+\,\spz{j+1}\,)\,)\nonumber\\
 & &\mbox{} +\frac{1}{2}\, (1\,-\,
\spx{1}\,\spx{2}\,)\,(1-\frac{1}{2}\,
 (\,\spz{1}\,+\,\spz{2}\,)\,)\nonumber\\
 & &\mbox{} + \frac{1}{2}\,(1\,\mp\,\
\hat{Q}_{L}\,\spx{L}\,\spx{1}\,)
\,(1-\frac{1}{2}\,
 (\,\spz{1}\,+\,\spz{L}\,)\,)\enspace . 
\label{9}
\end{eqnarray}

 The operators $\tilde{H}^{\pm}$, 
 restricted to the subspaces $\hat{Q}_{L}=1$
 for $\tilde{H}^{+}$ and
 $\hat{Q}_{L}=-1$ for $\tilde{H}^{-}$,
 are equivalent to
 the Hamiltonian of the
 diffusion-annihilation
 problem with rates of 
 diffusion $1/2$ and rate of
 annihilation $1$ and periodic
 boundary conditions, which can
 be solved in terms of free fermions
 \cite{Alcaraz}. The other cases correspond
 to non-stochastic processes.
 Hence, we obtain
 a rigorous formulation of the well
 known duality transformation
 between the two models \cite{Racz}.
 These results are summarized in 
 Table 1. Notice that, although we
 have not used it explicitly, the similarity
 transformation preserves the relations
 (\ref{2}) 
 so one can also represent $\tilde{H}^{\pm}$
 in terms of Hecke algebra generators. However,
 one has used here an Hermitian quotient of the
 algebra which is different from the one
 used in references \cite{Alcaraz,Alcaraz2}
 and which allows the Hamiltonian for the diffusion-annihilation
 to be written as a linear combination of Hecke algebra
 generators.

\begin{table}[t]
\begin{center}
\begin{tabular}{|c|c|c|c|} \hline
 \multicolumn{2}{|c|}{Glauber Dynamics}& 
\multicolumn{2}{c|}{Annihilation Dynamics} \\
\hline
 Operators & Boundary Conditions & Operators
 & Boundary Conditions\\\hline
 $H^{+}$, $\hat{C}\;=\;1$ & Periodic& $\tilde{H}^{+}$, 
 $\hat{Q}_{L}\;=\;1$& Periodic\\
 $H^{-}$, $\hat{C}\;=\;1$ & Antiperiodic& $\tilde{H}^{-}$, 
 $\hat{Q}_{L}\;=\;-1$& Periodic\\
 $H^{+}$, $\hat{C}\;=\;-1$ & Antiperiodic&$\tilde{H}^{+}$, 
 $\hat{Q}_{L}\;=\;-1$ & Non physical\\
 $H^{-}$, $\hat{C}\;=\;-1$ & Periodic &$\tilde{H}^{-}$, 
 $\hat{Q}_{L}\;=\;1$ & Non physical\\
\hline
\end{tabular}
\caption{Summary of the relations between Glauber
 and annihilation dynamics. On the left hand side
 we have the Hamiltonian operator which is equivalent
 to the Glauber-Ising Hamiltonian and the sector
 of the Hilbert space where that equivalence holds,
 indicated by the eigenvalue of $\hat{C}$. On the right
 hand side we have the 
 tranformed Hamiltonian and the sector
 in which it is equivalent to the 
 diffusion-annihilation process
(given by the eigenvalue of $\hat{Q}_{L}$).}
\end{center}
\end{table}

 Finally, let us consider the
 action of $V_{\pm}$ in a single
 $\spz{j}$ operator. Expressing the operators
 $g_{l}$ in terms of Pauli spin matrices in
 (\ref{3}) one can, using the commutation
 relations for these operators show that
\begin{equation}
 V_{\pm}\,\spz{j}\,V_{\pm}^{-1}\;=\; - 
 \spy{1}\,\spz{2}\ldots\spz{j}\enspace.
 \label{11}
\end{equation}
  Using (\ref{11}), one obtains
 the following transformation
 law for a pair of 
 $\spz{k}\,\spz{l}\;\;k<l$
\begin{eqnarray}
 V_{\pm}\,\spz{k}\,
 \spz{l}\,V_{\pm}^{-1}&=&
 (V_{\pm}\,\spz{k}\,V_{\pm}^{-1})
 \,(V_{\pm}\,\spz{l}\,
 V_{\pm}^{-1})\nonumber\\
 &=&(\spy{1}\,\spz{2}
\ldots\spz{k})\,(\spy{1}\,
\spz{2}\ldots\spz{k}\ldots\spz{l})
\;=\;\spz{k+1}\ldots\spz{l}\enspace.
\label{12}
\end{eqnarray}
 Note that taking $l=k+1$ or
 $k=L$, $l=1$ we recover the 
 equalities (\ref{8}) concerning
 $\spz{k+1}$. Equation (\ref{12})
 will be useful below when we
 derive equalities concerning correlation
 functions. We now proceed to study
 the effect of $V_{\pm}$ in the states
 of the theory.

\section{The transformation law for the states}
 The Glauber-Ising Hamiltonian at T$=0$
 has two ground states, the ferromagnetic
 states with all
 spins up or down. Since $H^{+}$ is only
 equivalent to it
 in the subspace of the states
 with $\hat{C}\,=\,1$
 we have to find a linear
 combination of these two states
 that belongs to this subspace
(for simplicity we will specialize in $V_{+}$).
 This state is simply $\ket{\Psi}\,=
 \,\frac{1}{2}\,(\ket{\uparrow\ldots
 \uparrow}\;+\;\ket{\downarrow
 \ldots\downarrow})$. It is
 normalized in the sense that
 $\bra{s}\,\Psi\,\rangle\,=\,1$,
 where $\bra{s}$ is the 
 sum of all configurations 
 of spins with weight one.
 This expresses the
 fact that, for a stochastic process,
 the sum of the probabilities
 of the different configurations
 accessible to the system (e.g. the
 configurations of spins
 in the Glauber problem) has
 to add to one.
 Also if $H$ is the stochastic Hamiltonian
 representing the dynamics then it
 follows from conservation of probability
 that $\bra{s}$ is the left ground state
 of $H$, i.e. $\bra{s}\,H=0$.
 One can easily check that
 $\hat{C}\,\ket{\Psi}\,=\,
\ket{\Psi}$. This equality implies
 that 
\beq
 V_{+}\,\ket{\Psi}\;=\;V_{+}
 \,\hat{C}\,\ket{\Psi}\;=\; 
 (V_{+}\,\hat{C}\,V_{+}^{-1})
\,V_{+}\,\ket{\Psi}\;=\;
\hat{Q}_{L}\,V_{+}\,\ket{\Psi}
\label{13}
\eeq
 where we have used the
 transformation law for $\hat{C}$
 found above.
 Thus, the transformed state
 $V_{+}\,\ket{\Psi}$ belongs
 to the eigenspace with $\hat{Q}_{L}=1$.
 In the annihilation
 language this corresponds to the
 sector with an even number
 of particles \cite{Schutz1}. Also,
 since $H^{+}\,\ket{\Psi}\,=\,0$,
 one obtains
\beq
 V_{+}\,H^{+}\,\ket{\Psi}\;=\; 
 (V_{+}\,H^{+}\,V_{+}^{-1})
 \,V_{+}\,\ket{\Psi}\;=\;
 \tilde{H}^{+}\,V_{+}\,\ket{\Psi}\;=\;0
\label{14}
\eeq
 and hence $V_{+}\,\ket{\Psi}$
 is a ground state of the
 annihilation Hamiltonian (in the
 subspace $\hat{Q}_{L}=1$,
 $\tilde{H}^{+}$ is equivalent to it).
 The only ground state belonging
 to the subspace with an even
 number of particles is the
 vacuum $\ket{0}$, i.e. the
 state with no particles.
 Hence, we conclude that
 $V_{+}\,\ket{\Psi}\,\propto\,\ket{0}$.
 The proportionality constant can be
 shown, using (\ref{3}), to be equal to 
 $\frac{i}{\sqrt{2}}(-1)^{L-1}e^{i\frac{\pi}{4}(L-1)}$
 and can be absorbed in the definition of $V_{+}$.
 Following the same reasoning
 that led to (\ref{14}), and
 in light that $\bra{s}\,\hat{C}\,=\,
 \bra{s}$,
 one can similarly show
 that $\bra{s}\,V_{+}^{-1}\,
 \tilde{H}^{+}\,=\,0$
 and that $\bra{s}\,V_{+}^{-1}$
 belongs to the even sector.
 On the same grounds
 of uniqueness this shows
 that this state is equal
 (up to a normalisation constant
 which we absorb in the definition of
 $V_{+}^{-1}$)
 to the left ground state
 of the annihilation Hamiltonian
 with an even number of particles, i.e.
 $\bra{s}^{even}$, which is
 the sum of all configurations
 with an even number of particles
 with weight one.
 We are using the same notation
 $\bra{s}$ for the left
 ground states of the two
 Hamiltonians because both
 describe stochastic processes
 and this is the usual
 convention.

 We have also found 
 that $H^{-}$ is equivalent
 to the Glauber-Ising Hamiltonian
 in the subspace $\hat{C}\,=\,-1$.
 Since  $V_{-}\,\hat{C}
 \,V_{-}^{-1}\,=\,-\,\hat{Q}_{L}$
 this subspace is mapped to the
 even sector of the annihilation
 problem. But in this sector
 $\tilde{H}^{-}$ (\ref{9}) is not
 equivalent to a stochastic Hamiltonian.
 If one starts with the subspace
 $\hat{C}\,=\,1$ then $V_{-}$ effectively
  maps this sector to the odd
 sector where $\tilde{H}^{-}$ is
 equivalent to the annihilation Hamiltonian,
 but in this case $H^{-}$ is not
 equivalent to the Glauber-Ising Hamiltonian
 with periodic boundary conditions, but to
 the Glauber-Ising Hamiltonian with
 anti-periodic boundary conditions,
 which is also a stochastic process.
 Using the same argument as above one
 can show that the ground state of this
 Hamiltonian is mapped to the ground state
 of the odd sector of the annihilation Hamiltonian.
 This state is just a uniform superposition of
 states with one particle at each site of
 the lattice \cite{Schutz1}. If we apply $V_{-}^{-1}$
 to this state we will therefore obtain the ground
 state of the Glauber-Ising Hamiltonian with
 anti-periodic boundary conditions (the 
 observation about normalisation factors
 also applies here). Such 
 state is a uniform superposition of
 $2L-2$ states with two domain walls, one
 at the boundary and one at each site of the
 lattice, plus the two states with all
 spins up or down. These two states are,
 due to the anti-periodic boundary conditions,
 the image of the state with one particle
 at the boundary in the diffusion-annihilation
 model. 
 The previous discussion shows 
 that the transformations
 $V_{+}$  and $V_{-}$
 restricted to the $\hat{C}=1$
 subspace, map a stochastic
 process (Glauber-Ising dynamics
 with periodic or anti-periodic boundary
 conditions) to
 a stochastic process (annihilation
 dynamics), something which we also
 pointed out in Table 1. In the other cases
 the mapping is from a stochastic
 problem to a non-stochastic problem. 
 Nevertheless, these mappings can be considered 
 from a purely formal point of view and examples
 of similarity transformations  between stochastic
 and non-stochastic processes have
 already been studied in
 the literature \cite{Henkel,Krebs,Simon}. 

 Here we shall concentrate in the
 stochastic-stochastic mapping
 given by $V_{+}$. We
 will consider the study of time dependent
 correlation functions in uncorrelated
 random initial states evolving in time
 according to Glauber
 dynamics. The state
\beq
\ket{\Phi}\;=\;\prod_{j=1}^{L}\left[\,
\frac{1+m}{2}\;+\;\frac{1-m}{2}\,
\spx{j}\,\right]\,\ket{\Psi}
\label{15}
\eeq
 corresponds to the superposition
 of two random initial states
 with initial magnetization $m$
 and $-m$. From the point of view
 of the calculation of correlation
 functions of an even number of
 $\spz{j}$ operators the two states
 are equivalent and $\ket{\Phi}$
 belongs to the $\hat{C}\,=\,1$
 subspace. Under the application of
$V_{+}$, $\ket{\Phi}$ transforms to
\beq
\ket{\tilde{\Phi}}\;=\;
\prod_{j=1}^{L}\left[\,
\frac{1+m}{2}\;+\;\frac{1-m}{2}
\,\spx{j}\,\spx{j+1}\,\right]\,\ket{0}
\label{16}
\eeq
 where we have used (\ref{8})
 and the fact that
 $V_{+}\,\ket{\Psi}\,=\,\ket{0}$.
 This is an initial
 state with short range correlations
 and its form will play
 a crucial role in the
 determination of the correlation
 functions in the late time
 regime. Under $V_{+}$ the multiple
 time  correlation
 functions of an even
 number of $\spz{j}$ spins at
 times $t_{1}$, $t_{2}$,
 etc., transform as
\begin{eqnarray}
& &\bra{s}\,\spz{j_{1}}(t_{1})
\,\spz{j_{2}}(t_{1})\,
\spz{j_{3}}(t_{2})\,
\spz{j_{4}}(t_{2})\ldots
\spz{j_{2N-1}}(t_{N})\,
\spz{j_{2N}}(t_{N})\,
\ket{\Phi}\nonumber\\ 
 &=&\bra{s}\,\spz{j_{1}}\,\spz{j_{2}}
\,e^{-H^{+}(t_{1}-t_{2})}
\spz{j_{3}}\,
\spz{j_{4}}\,e^{-H^{+}(t_{2}-t_{3})}\ldots 
 e^{-H^{+}(t_{N-1}-t_{N})}\,
\spz{j_{2N-1}}
\,\spz{j_{2N}}\,e^{-H^{+}t_{N}}
\ket{\Phi}\nonumber\\ 
&=&\bra{s}^{even}\,\spz{j_{1}+1}
\ldots\spz{j_{2}}
\,e^{-\tilde{H}^{+}(t_{1}-t_{2})}
\spz{j_{3}+1}\ldots
\spz{j_{4}}\,e^{-\tilde{H}^{+}
 (t_{2}-t_{3})}\nonumber\\ 
& &\mbox{}\times
\ldots e^{-\tilde{H}^{+}
 (t_{N-1}-t_{N})}\,
\spz{j_{2N-1}+1}
\ldots\spz{j_{2N}}\,
 e^{-\tilde{H}^{+}t_{N}}
\ket{\tilde{\Phi}}\nonumber\\ 
 &=&\bra{\!s}\spz{j_{1}+1}(t_{1})
\ldots\spz{j_{2}}(t_{1})
\spz{j_{3}+1}(t_{2})\ldots
\spz{j_{4}}(t_{2})\ldots
\spz{j_{2N-1}+1}(t_{N})
\ldots\spz{j_{2N}}(t_{N})
\ket{\tilde{\!\Phi}}
\label{17}
\end{eqnarray}
 where we have used
 equations (\ref{9}) and (\ref{12}).
 We suppose that $j_{1}<j_{2}$,
 $j_{3}<j_{4}$, etc.
 Also one concludes that for any odd
 number of $\spz{j}$ operators, one has
\beq
\bra{s}\,\spz{j_{1}}(t_{1})\ldots
\spz{j_{2N+1}}(t_{2N+1})\,\ket{\Phi}\;=\;0,
\label{18}
\eeq
 since $\spz{j}$ anticommutes with $\hat{C}$.
\section{The calculation of correlation functions}
 Using (\ref{17}) we have, in particular, that
\beq
\bra{s}\,\frac{1}{2}\,(1\,-\,
\spz{j-1}(t)\,\spz{j}(t)\,)\,\ket{\Phi}\;=\;
\bra{s}\,\frac{1}{2}\,(1\,-\,
\spz{j}(t)\,)\,\ket{\tilde{\Phi}}\;=\;
\bra{s}\,\noc{j}(t)\,
\ket{\tilde{\Phi}}\enspace.
\label{19}
\eeq
 The operator $\frac{1}{2}\,(1\,-\,
 \spz{j-1}\,\spz{j})$ checks
 for the existence of a domain
 wall at site $j$ in the Glauber problem,
 i.e. it checks if at site $j$
 the spins cease to point up and start
 to point down or vice-versa.
 The operator 
 $\frac{1}{2}\,(1\,-\,\spz{j}\,)$
 checks for the existence of a particle
 at site $j$, 
 i.e. if spin $j$ is down in
 the annihilation problem \cite{Schutz1}.
 This indeed corresponds to the
 well known duality transformation
 \cite{Racz,Family}. Similar relations
 hold for higher order correlation 
 functions. Now we will use the fact
 that the diffusion annihilation 
 Hamiltonian can be completely
 solved in terms of free fermions
 by means of the Jordan-Wigner
 transformation \cite{Jordan}. Using it,
 one is able to write the spin
 raising and lowering operators
 $\spup{j}$, $\spdo{j}$ ($
\spupdo{j}\,=\,\frac{1}{2}\,(\spx{j}
\,\pm\mbox{i}\,
\spy{j}\,)$) in terms of creation
 and annihilation operators 
 $\crea{a}{j}$, $\anni{a}{j}$ for
 spinless fermions.
 It turns out that the calculation
 of a multiple time correlation
 function of the $\noc{j}$ operators
 in state $\ket{\tilde{\Phi}}$
 ($\noc{j}\,=\,\crea{a}{j}\,
\anni{a}{j}$ in the fermion language)
 can be reduced to the
 calculation of objects like
 $\bra{s}\,\anni{b}{p_{1}}\ldots
\anni{b}{p_{2k}}\,\ket{\tilde{\Phi}}$
 ($k\leq N$) \cite{Santos} where $\anni{b}{p}$ is
 the Fourier transform of the annihilation
 operator $\anni{a}{i}$. The
 momentum labels $p_{j}$ are half-odd
 integers between $-L/2 +1$
 and $L/2$ \cite{Alcaraz,Schutz1}. Also, one
 is able to express the state (\ref{16})
 in terms of fermion operators. The result is
\begin{eqnarray}
\ket{\tilde{\Phi}}&=&\prod_{j=1}^{L}\left[\,
\frac{1+m}{2}\;+\;\frac{1-m}{2}\,\spx{j}
\,\spx{j+1}\,\right]\,\ket{0}
\;=\;\exp\left(\beta\sum_{j=1}^{L}\,(\,\spx{j}\spx{j+1}
\,-\,\mbox{\small $1$}\,)\right)\ket{0}\nonumber\\
 &=&\exp\left(\beta\sum_{j=1}^{L}\,(\,
\crea{a}{j}\,\crea{a}{j+1}\,+\,\crea{a}{j}\,
\anni{a}{j+1}\,+\,\crea{a}{j+1}\,\anni{a}{j}\,+
 \,\anni{a}{j+1}\,\anni{a}{j}\,-\,
\mbox{\small $1$}\,)\right)\,\ket{0}
\label{20}
\end{eqnarray}
 where $m\;=\;e^{-2\beta}$. The first equality
 follows from the fact that 
 $(\,\spx{j}\,\spx{j+1}\,)^{2}\,=\,1$
 and the second just follows from
 the rules of the Jordan-Wigner
 transformation. In terms of
 the momentum space operators
 we can write $\ket{\tilde{\Phi}}$ as
\beq
\ket{\tilde{\Phi}}=\exp\left(2\beta\sum_{p>0}\,
 [\,\cos\left(\frac{2\pi p}{L}\right)
\,(\,\crea{b}{p}\,\anni{b}{p}\,+\,
\crea{b}{-p}\,\anni{b}{-p}\,)\,+\,
\sin\left(\frac{2\pi p}{L}\right)
\,(\,\anni{b}{p}\,\anni{b}{-p}\,+\,
\crea{b}{-p}\,\crea{b}{p}\,)\,-\,
\mbox{\small $1$}\,]\right)\,\ket{0}\enspace.
\label{21}
\eeq
 The simplest approach to use if one wants to calculate
 the density or any equal-time correlators is given in
 \cite{Schutz1}. For the calculation of multiple time
 correlators, we will follow a different route.
 We will look for operators 
 $\crea{c}{p}$, $\anni{c}{p}$ 
 that diagonalize the quadratic
 form appearing in the exponent of 
 (\ref{21}). This can be accomplished
 by means of a Bogoliubov
 transformation \cite{Bogoliubov}
\begin{equation}
\left\{
\begin{array}{l}
\anni{c}{p}\;=\;\cos\left(\frac{
\pi p}{L}\right)\,\anni{b}{p}\;-\;
\sin\left(\frac{\pi p}{L}\right)
\,\crea{b}{-p}\\ \\
\crea{c}{p}\;=\;\cos\left(\frac{
\pi p}{L}\right)\,\crea{b}{p}\;-\;
\sin\left(\frac{\pi p}{L}\right)\,\anni{b}{-p}
\end{array}
\right.
\label{22}
\end{equation}
 and one gets
\beq
\ket{\tilde{\Phi}}\;=\;\exp\left(2\beta\sum_{p}
\,(\,\crea{c}{p}\,\anni{c}{p}
\,-\,\mbox{\small $1$}\,)\right)\,\ket{0}\enspace.
\label{23}
\eeq

 One then expands the exponential
 using the anti-commutation relations 
  for the  $\crea{c}{p}$, $\anni{c}{p}$.
 Re-expressing these operators
 in terms of $\crea{b}{p}$,
 $\anni{b}{p}$ and applying them to
 the vacuum, one finally obtains
\beq
\ket{\tilde{\Phi}}\;=\;e^{-\beta L}
\prod_{p>0}\,[\,
\gamma_{p}\;+\;\delta_{p}\,\crea{b}{-p}\,
\crea{b}{p}\,]\,\ket{0}
\label{24}
\eeq
 where $\gamma_{p}\,=\,\cosh(2\beta)
\,-\,\sinh(2\beta)\,
\cos\left(\frac{2\pi p}{L}\right)$
 and $\delta_{p}\,=\,
\sinh(2\beta)\,
\sin\left(\frac{2\pi p}{L}\right)$.
 The form (\ref{24})
 expresses the translation
 invariance of the state
 $\ket{\tilde{\Phi}}$.
 One defines the following object
\beq
 Z[\eta_{p},\eta_{-p}]\;=\;\bra{0}\,
 \exp\left(\sum_{p>0}(\cot(\frac{
 \pi p}{L})\,\anni{b}{p}
 \,\anni{b}{-p}+\etali{p}\,
 \anni{b}{p}+\etali{-p}\,
 \anni{b}{-p})\right)
 \,\ket{\tilde{\Phi}}
\label{25}
\eeq
 where the quantities \(\eta_{p}\), \(\eta_{-p}\) 
 are Grassmann variables anticommuting among 
 themselves and with the \anni{b}{p}'s. Their 
 presence is necessary to make the terms in the 
 exponential commute with each other.
 It can be shown \cite{Santos}
 that $Z[\eta_{p},\eta_{-p}]$
 is the generating function for the quantities
 $\bra{s}\,\anni{b}{p_{1}}\ldots
\anni{b}{p_{2k}}\,\ket{\tilde{\Phi}}$
 that is
\beq
\bra{s}\,\anni{b}{p_{1}}\ldots
\anni{b}{p_{2k}}\,\ket{\tilde{\Phi}}\;=\;
\partial_{\eta_{p_{1}}}\ldots
\partial_{\eta_{p_{2k}}}
\,Z[\eta_{p},
\eta_{-p}]\mid_{\eta_{p}=0}\enspace.
\label{26}
\eeq

 Substituting (\ref{24}) in (\ref{25})
 we find, after
 a few algebraic manipulations,
 involving the anticommutation
 relations between the $\crea{b}{p}$,
 $\anni{b}{p}$ \cite{Santos},
 the following expression
 for $Z[\eta_{p},\eta_{-p}]$
\begin{eqnarray}
 Z[\eta_{p},\eta_{-p}]&=&
  \exp\left(\sum_{p>0}\,
\sinh(2\beta)\,e^{-2\beta}\sin\left(
\frac{2\pi p}{L}\right)
\,\eta_{-p}\eta_{p}\right)\enspace.
\label{27}
\end{eqnarray}

 Since $W\,=\,\ln\,Z$
 is a quadratic function
 in the $\eta$'s it
 immediately follows that
 a Wick's decomposition
 holds for the quantities
 $\bra{s}\,\anni{b}{p_{1}}\ldots
\anni{b}{p_{2k}}\,
\ket{\tilde{\Phi}}$ \cite{Santos}.
 In particular, from (\ref{26}) and
 (\ref{27}) one has
\beq
\bra{s}\,\anni{b}{p'}\,\anni{b}{p}\ket{
\tilde{\Phi}}\;=\;
\sinh(2\beta)\,e^{-2\beta}\sin\left(
\frac{2\pi p'}{L}\right)
\,\delta_{p,-p'}\enspace.
\label{28}
\eeq

 Given that the expression for the space-dependent
 average density is \cite{Schutz1}
\beq
\langle\noc{j}(t)\rangle\;=
\;\frac{1}{L}\,\sum_{p,p'}\,
 e^{\frac{2\pi i}{L}j(p-p')-(\epsilon_{-p}
 +\epsilon_{p'})t}\,\cot\,
\left(\,\frac{\pi p}{L}\,\right)\,
\bra{s}\,\anni{b}{p'}\,\anni{b}{-p}
\,\ket{\tilde{\Phi}}
\label{29}
\eeq
 where  \(\epsilon_{p}\,=\,1\,-\,\cos\,p\),
 one obtains, substituting
 (\ref{28}) in (\ref{29})
 and taking the thermodynamic
 limit $L\rightarrow\infty$,
 the following expression
 for the density of particles
 $\rho(t)$ at time $t$
\beq
\rho(t)\;=\;\frac{1}{2}
\,(1\,-\,m^{2})\,e^{-2t}
\,(\,I_{0}(2t)\;+\;I_{1}(2t)\,)
\label{30}
\eeq
 where $I_{0}(2t)$, $I_{1}(2t)$ are the modified
 Bessel functions of order zero and one, and where
 we have used the identity $m=e^{-2\beta}$.
 This is the well-known expression obtained by
 Family and Amar \cite{Family}
 who have also considered
 a random initial state
 in Glauber dynamics. They
 have shown that while
 the initial
 distribution of spins is
 uncorrelated the distribution
 of domain walls,
 i.e. the distribution
 of particles in the annihilation
 problem, is correlated.
 This correlated structure
 is found in the transformed state
 $\ket{\tilde{\Phi}}$ (\ref{16}). For $m=0$, 
 the expression
(\ref{30}) is identical to the one found by
 Spouge \cite{Spouge} for an
 uncorrelated random initial 
 state with initial density $1/2$
 in the annihilation
 problem. Indeed, this
 identity is more general.
 If $m=0$ then this means
 that we have to take
 the limit $\beta\rightarrow
 \infty$. If we take
 such a limit in equation
 (\ref{24}) then one obtains
 the exact expression for a
 random initial state with
 density $1/2$, projected
 over the even sector \cite{Schutz1}.
 Therefore, our calculations
 provide a rigorous
 framework for the well-known
 duality transformation.
 In the long time limit
 $t\rightarrow\infty$ one 
 finds from (\ref{30}) the leading behaviour
 $\rho(t)\,\approx\,\frac{1-m^{2}}{2\sqrt{\pi t}}$.
 It depends on the initial conditions
 (i.e. magnetization) \cite{Family}.
 The amplitude of $\rho(t)$
 differs from
 the universal result found
 for uncorrelated random
 initial states \cite{Torney,Krebs}.
 Thus it is seen that
 the presence of correlations
 in the initial state
 breaks the universality of the 
 amplitudes of correlation functions.

 Our results, namely the Wick
 decomposition, also
 allow the explicit computation of
 higher order correlation
 functions. Apart from a factor
 $2\sinh(2\beta)\,e^{-2\beta}$
 the result (\ref{28}) is identical
 to the results obtained
 for the random initial state
 with density $1/2$ \cite{Santos} and,
 as stated above, it reduces to
 it when $m=0$. Therefore one has
\beq
\bra{s}\,\anni{b}{p_{1}}\ldots
\anni{b}{p_{2k}}\,\ket{\tilde{\Phi}}\;=\;
 (\,1\,-\,e^{-4\beta}\,)^{k}\,
\bra{s}\,\anni{b}{p_{1}}\ldots
\anni{b}{p_{2k}}\,\ket{1/2}^{even}\enspace.
\label{31}
\eeq

 As an example of the above
 result let us consider the
 two-point correlation function 
 $\bra{s}\,\noc{j}(t)\,\noc{k}(t')
 \ket{\tilde{\Phi}}$.
 We use (\ref{31}) with $k=2$
 and $k=1$ because
 this object can be written
 as a linear combination of 
 the terms $\bra{s}\,\anni{b}{p_{1}}
\,\anni{b}{p_{2}}\,
\anni{b}{p_{3}}\anni{b}{p_{4}}
\,\ket{\tilde{\Phi}}$
 and  $\bra{s}\,
\anni{b}{p_{1}}\,\anni{b}{p_{2}}\,
\ket{\tilde{\Phi}}$ \cite{Santos}, which
 contribute with different
 powers of $(\,1\,-\,e^{-4\beta}\,)$.
 Separating such powers, we obtain
\begin{equation}
\left.
\begin{array}{l}
\bra{s}\,\noc{j}(t)\,
\noc{k}(t')\ket{\tilde{\Phi}}=
 (\,1\,-\,e^{-4\beta}\,)^{2}\,
\bra{s}\,\noc{j}(t)\,
\noc{k}(t')\ket{1/2}^{even}\\ 
\mbox{}+\frac{1}{2L^{2}}\,
 [\,(1-e^{-4\beta})-
(1-e^{-4\beta})^{2}]\;\times\\ 
\left\{\,\sum_{p_{1},p_{2}}
 e^{\frac{2\pi i}{L}(p_{1}-p_{2})
(j-k)-(\epsilon_{-p_{1}}
 +\epsilon_{p_{2}})t+(\epsilon_{-p_{1}}-
\epsilon_{-p_{2}})t'}
\!\sin\left(\frac{2\pi p_{2}}{L}\right)
\cot\left(\frac{
\pi p_{1}}{L}\right)\right.\\ 
\left.\mbox{}+\,\sum_{p_{1},p_{2}}
 e^{\frac{2\pi i}{L}(p_{1}-p_{2})
 (j-k)-(\epsilon_{-p_{1}}
 +\epsilon_{p_{2}})t-
 (\epsilon_{p_{1}}-\epsilon_{p_{2}})t'}
\!\sin\left(\frac{2\pi p_{1}}{L}\right)
\cot\left(\frac{\pi p_{1}}{L}
\right)\right\}\enspace.
\end{array}
\right.
\label{32}
\end{equation}

 In the thermodynamic limit
 $L\rightarrow\infty$, we
 have for $t'=t$
\begin{eqnarray}
\bra{s}\noc{j}(t)\,\noc{k}(t)
\ket{\tilde{\Phi}}\!\!\!&=&\!\!\!
(1-e^{-4\beta})^{2}\,
\bra{s}\,\noc{j}(t)\,\noc{k}(t)
\ket{1/2}^{even}\nonumber \\
 & &\!\!\!\mbox{}+ e^{-4\beta}
\,(1-e^{-4\beta})\times
\left\{\frac{1}{2}e^{-2t}(\,I_{0}(2t)\,+
\,I_{1}(2t)\,)\,\delta_{j,k}\right.\nonumber \\
 & &\left.\!\!\!\mbox{}+\!
\frac{1}{4}[\,\theta(j-k)-\theta(k-j)\,]\,e^{-2t}\,
(\,I_{j-k-1}(2t)-I_{j-k+1}(2t)\,)\right\}
\label{33}
\end{eqnarray}
 where we have used the
 integral representation of the
 modified Bessel functions $I_{j}(2t)$
 and $\theta(x)$
 is the Heaviside step function.
 The term $\theta(j-k)-\theta(k-j)$
 is the Fourier transform of 
 $\cot\left(\frac{\pi p}{L}\right)$ 
 \cite{Santos}.
 In particular, when $t=0$,
 this reduces to
\begin{eqnarray}
\bra{s}\noc{j}\,\noc{k}
\ket{\tilde{\Phi}}\!\!\!&=&\!\!\!
\frac{1}{2}(1-e^{-4\beta})
\,\delta_{j,k}\;+\;
\frac{1}{4}(1-e^{-4\beta})^{2}
\,(1-\delta_{j,k})\nonumber \\
 & &\!\!\!\mbox{}+\frac{1}{4}
\,e^{-4\beta}(1-e^{-4\beta})
\,(\,\delta_{j,k+1}\,+\,\delta_{j,k-1}\,)
\label{34}
\end{eqnarray}
 where we have used
 the fact that there are
 no correlations in the
 state $\ket{1/2}^{even}$
 at $t=0$. One clearly
 recognizes in the first
 two terms the contribution
 of the unconnected
 part of the correlation function.
 But the third term
 indeed confirms that even
 at $t=0$ there are short
 range correlations. This term is zero
 when we take $\beta
 \rightarrow\infty$ and 
 we just obtain the trivial result 
 for the $\ket{1/2}^{even}$ state.

 If we use equation 
 (\ref{32}) to calculate
 the density-density correlation 
 function in the thermodynamic limit
 we see that the second
 term of (\ref{32}) will
 vanish. This leaves us with a
 term with an amplitude proportional to
 $(1-m^{2})^{2}$. The ratio
 of this correlation function
 with the square
 of the density (\ref{30}) is
 independent of $m$ in agreement with
 the general results
 known from the renormalisation
 group approach
 (see for example \cite{Cardy}).
 So despite the fact
 that the amplitudes of
 the various correlation
 functions are non-universal
 as emphasized above, their
 ratios obey
 the universality hypothesis.

 The results discussed
 above show that this
 approach allows
 not only to recover the
 know results but also
 provides a convenient way to
 compute higher order correlation
 functions that can of course
 be translated back to the
 Glauber-Ising language.

\section{Conclusions}
 We investigated the relation
 between the Glauber-Ising
 model at zero temperature and
 the diffusion-anni\-hi\-la\-tion
 model in the free fermion case.
 We obtained the following new
 results

 (i) The duality transformation
 between the two models
 can be formulated as a similarity
 transformation if one
 uses an Hamiltonian with
 sector-dependent toroidal
 boundary conditions. The
 transformation laws for the
 operators are
 explicitly given. We also
 obtain the transformation
 laws for the states. This
 permits a {\em one-to-one}
 correspondence between
 a state of Glauber-Ising
 dynamics and a state of
 the diffusion-annihilation
 problem. In particular
 an uncorrelated  random
 initial state in Glauber dynamics
 transforms to a state
 with short range correlations.

 (ii) Using the free fermion
 solution of the
 diffusion-annihilation
 problem we have
 computed the time dependent 
 behaviour of the density
 and equal time two point
 correlation function in this
 short-range correlated state.
 For the density, we
 recover the 
 results of the literature.
 We show that, surprisingly, the
 presence of correlations 
 extending over only one
 lattice site in the
 initial state leads
 to a long time behaviour
 of the density dependent
 on the initial condition.
 The field theoretic
 approach we have used can
 be applied to the study of
 higher order correlation
 functions. Its use depends 
 on the form (\ref{24}) 
 of the initial
 state which reflects the
 translation invariance
 of this state.
 An initial state with the Wick's
 decomposition property was
 also discussed by Balboni et al.
 \cite{Balboni}. They have considered
 a continuous system described
 by a boson field theory. The initial state
 which they have chosen is characterized
 by the fact that the connected
 correlation functions of the
 density operator of 
 order higher than two vanish. They
 found that for pure annihilation
 the amplitudes are also non-universal.
 In the case of the initial state (\ref{16})
 the higher order connected
 correlation functions of
 the density operator
 are non-zero, which shows that this state 
 has a different structure. 
\\ \\
\noindent{
{\bf Acknowledgments:}
 I am grateful to G. Sch\"{u}tz
 for having introduced me to the concept
 of duality transformation and
 associated Temperley-Lieb
 and Hecke algebras and for carefully reading
 the manuscript.
 It is also a pleasure to acknowledge
 many fruitful discussions
 with Robin Stinchcombe,
 Michel Droz, Zoltan Racz,
 Haye Hinrichsen and John Cardy.
 I also thank Tim Newman
 for having called my attention
 to the universality of
 the ratios of amplitudes of
 correlation functions. The author
 is  supported by the
 Grant: PRAXIS XXI/BD$/3733/94$ - JNICT -
 PORTUGAL.}

\pagebreak

\end{document}